\documentclass[twocolumn,preprintnumbers,amsmath,amssymb]{revtex4-1}

\usepackage{array}
\usepackage[dvipdfmx]{graphicx}
\usepackage{tabularx}
\usepackage{longtable}
\usepackage{dcolumn} 
\usepackage{bm}
\usepackage{color}
\usepackage{braket}
\usepackage{mathtools}
\usepackage[utf8]{inputenc}

\begin{document}

\title{Thermopower in transition-metal perovskites}

\author{Wataru Kobayashi}
\email{kobayashi.wataru.gf@u.tsukuba.ac.jp}
\affiliation
{Division of Physics, Faculty of Pure and Applied Sciences, University of Tsukuba, 
Ibaraki 305-8571, Japan}
\affiliation
{Tsukuba Research Center for Energy Materials Science {\rm (TREMS)}, 
University of Tsukuba, Ibaraki 305-8571, Japan}

\begin{abstract}

High-temperature thermopower is interpreted as entropy that a carrier carries. 
Owing to spin and orbital degrees of freedom, a transition metal perovskite exhibits large thermopower 
at high temperatures. 
In this paper, we revisit the high-temperature thermopower in the perovskites to 
shed light on the degrees of freedom. Thus, we theoretically derive an expression of thermopower 
in one-dimensional octahedral-MX$_6$-clusters chain using linear-response theory 
and electronic structure calculation of the chain based on the tight-binding approximation. 
The derived expression of the thermopower is consistent with the extended Heikes formula and 
well reproduced experimental data of several perovskite oxides at high temperatures. 
In this expression, a degeneracy of many electron states in octahedral ligand field (which is 
characterized by multiplet term) appears instead of the spin and orbital degeneracies. 
Complementarity in between our expression and the extended Heikes formula is discussed.

\end{abstract}

\maketitle
\section{Introduction}

Transition-metal (M) oxides are one of fascinating quantum many-body systems, which exhibits 
Mott insulating state \cite{mott}, spin-state crossover \cite{goodenough}, peculiar magnetism 
accompanied by orbital ordering \cite{kugel}, 
high-temperature superconductivity \cite{bednorz}, 
huge magneto-resistance \cite{tokura}, large thermopower \cite{terasaki}, 
exotic superconductivity \cite{maeno,takada,kamihara}, multiferroicity \cite{kimura}, 
spin-orbit related Mott insulating state \cite{kim}, spin frustration (possibly spin liquid) \cite{kitagawa}, 
and other relativistic effects such as Weyl semi metallic state \cite{ohtsuki}. 
A model material of the transition-metal oxides is a 3d-transition-metal perovskite oxide 
expressed as Ln$_{1-x}$Ae$_x$MO$_3$ ($0\le x \le 1$, Ln: lanthanide, Ae: alkali earth metal, and O: oxygen.). 
This system has been extensively studied. In particular, metal-insulator transition (MIT) was 
deeply understood in terms of filling, band-width, and dimensional controls \cite{imada}. 
The filling control is given by a substitution of Ae$^{2+}$ ion 
for Ln$^{3+}$ ion (hole doping). 
The band width ($W$) is controlled by ionic radius ($r_{\rm A}$) of element at A site. 
Smaller $r_{\rm A}$ causes smaller M-O-M bond angle, which realizes smaller $W$. 
Tolerance factor ($\tau$) is a function of $r_{\rm A}$ and a measure of strain in the perovskite 
structures. For $\tau \sim 0.9-1.0$, cubic structure with M-O-M bond 
angle of 180$^{\circ}$ realizes. 
With decreasing $r_{\rm A}$ as La$^{3+}$$\to$Pr$^{3+}$$\to$$\cdots$$\to$Y$^{3+}$$\to$Lu$^{3+}$, 
and Ba$^{2+}$$\to$Sr$^{2+}$$\to$Ca$^{2+}$, $\tau$ decreases which causes decrease in $W$. 
$\frac{U}{W}$ ($U:$ on-site coulomb interaction) is a key parameter, which gives 
a quantum critical point of the metal-insulator transition. 
Revealing electronic structures in the 3d-transition-metal perovskite oxides is generally difficult 
because of the electron correlation. Hubbard model \cite{hubbard} is a well-known model 
that duplicates the MIT in the strongly correlated systems \cite{imada,geroges}. 

Thermopower ($S$) is a phenomenological transport coefficient, and is defined as a temperature ($T$) 
derivative of thermoelectric voltage ($V$) as 
$S \equiv -\frac{dV}{dT}$, and can be applied for thermoelectric energy conversion \cite{mahan}. 
Thermopower at high-temperature limit in the quantum many-body systems 
is interpreted as entropy that a carrier carries in a regime of 
linear response theory with the Hubbard model \cite{heikes,chaikin}. Chaikin and Beni generalized 
Heikes formula \cite{heikes} to explain thermopower of interacting Fermi systems with spin, 
and explained thermopower of strongly correlated p-electron system 
(one-dimensional organic conductor) with spin entropy term 
$-\frac{k_{\rm B}}{e} \ln2$ \cite{chaikin}. 
Further, Doumerc extended the formula to apply for a system containing 
a mixed valent cation M$^{n+}$/M$^{(n+1)+}$ 
using spin multiplicity ($2S_n+1$) \cite{doumerc}. 
Marsh and Parris again extended the formula to take into account the orbital degeneracy, and 
qualitatively explained thermopower of LaCrO$_3$\cite{marsh1}, LaMnO$_3$\cite{marsh2} 
and their related perovskite systems. 
Koshibae {\it et al.} also explained the large thermopower of NaCo$_2$O$_4$ using 
the spin and orbital degrees of freedom and proposed application of their theory 
to Co perovskite oxides \cite{koshibae1}. 
In their theories, such degrees of freedom are introduced by counting a number of cases 
in many electrons 
configuration in e$_{\rm g}$ and t$_{\rm 2g}$ orbitals, and they seem to work well. 
To further consider the degrees of freedom, we 
derive a similar formula constructed from many electron states in MX$_6$ cluster with 
octahedral ligand (X: ligand element) field without a use of one-band Hubbard model. 

In this paper, we construct thermopower of one-dimensional MX$_6$-clusters chain model 
targeting the 3d-transition-metal perovskite oxides with $\frac{U}{W} \gg 1$. 
In a derived expression of the thermopower, 
we find that a degeneracy of many-electron states in the MX$_6$-cluster 
is included instead of the degeneracies of spin and orbital. 
We will discuss the relation between spin and orbital degrees of 
freedom and the degeneracy of the many-electron states. 
We further apply our formula to experimental data 
of several 3d-transition-metal perovskite oxides. Lastly, effects of 
spin-orbit interaction and Jahn-Teller distortion on the thermopower and further possible application 
of our theory will be discussed in 
the section of III. G.

\section{Methods}

In this section, first, we briefly review the crystal (ligand) field theory. 
We see many d-electron states in octahedral ligand field, term symbol, and 
how to count a degeneracy of the states that is represented by the term symbol. 
Second, we explain the tight-binding approximation (molecular-orbital theory) for the 
construction of the electronic structure of periodically aligned one-dimensional 
MX$_6$-clusters chain system (hereafter, clusters-chain model. See Fig. \ref{f2}). 
Third, we derive an expression of thermopower at high-temperature limit 
for the clusters-chain model, in a regime of linear-response theory. 
Last, we discuss complementarity in between the extended Heikes formula and our expression. 

\subsection{Crystal field theory (CFT)} 

\begin{figure}[t]
\begin{center}
\includegraphics[width=50mm,clip]{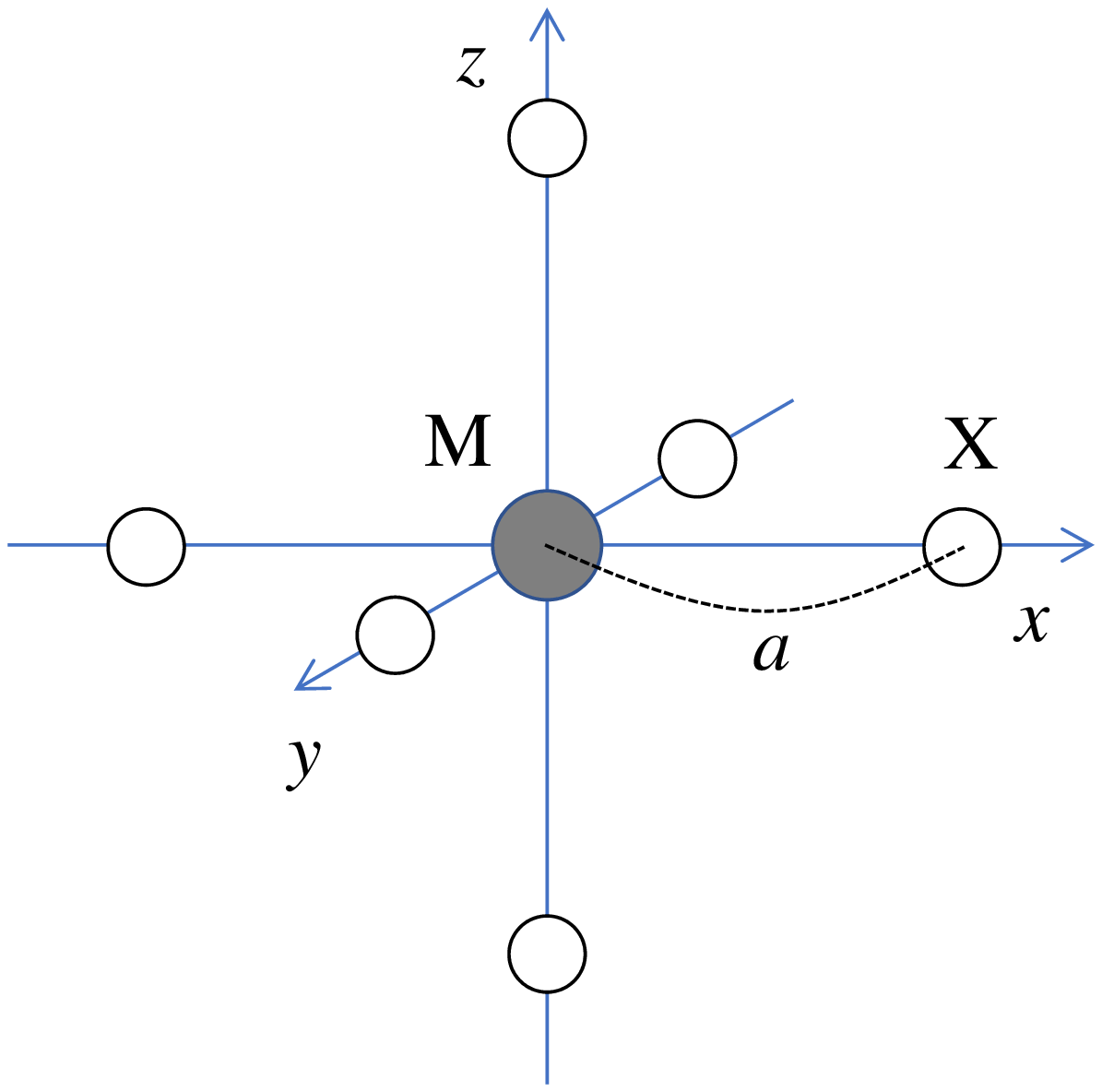}
\caption{(Color online) Schematic figure of MX$_6$ octahedron with $O_h$ symmetry. 
White circle represents a ligand (X) with its charge is $-Ze$ ($Z$: positive integer, $e$: 
charge unit). 
d electrons in 3d transition metal element (M) feel octahedral ligand field. 
}
\label{f1}
\end{center}
\end{figure} 

To calculate physical properties of solids, 
we generally need to treat its electronic structure correctly taking into account periodicity of the 
crystal. However, if on-site coulomb interaction ($U$) is much larger than inter-atomic transfer 
integral ($t$), we may discuss the physical properties of the crystal based on the atomic 
electronic states. Effects from other atoms can be treated as crystal field or 
ligand field which includes hybridization with p-orbitals of the ligands.   

Here, we briefly review the crystal field theory (CFT).  
CFT describes a breaking of a degeneracy of electron orbital states (d or f orbital states) 
of atom (or ion) due to a static electric field produced by a surrounding 
charge distribution. Suppose d-electrons of the atom (or the ion) are surrounded 
by 6 negative charges ($-Ze$, $Z$: positive integer, $e$: charge unit) at 
$\bm{R}_k=$ ($a$, 0, 0), (0, $a$, 0), (0, 0, $a$), ($-a$, 0, 0), (0, $-a$, 0), and (0, 0, $-a$).
Then, the Hamiltonian ($H_n$) of the $n$ d-electrons ($n$: a number of d-electrons) is described as 
\begin{equation}
H_n= \left(\sum_{i=1}^n \frac{\bm{p}_i^2}{2m_e}+v_{\rm core}(r_i)+v_{\rm CF}(r_i)\right)+\sum_{j>i=1}^n \frac{e^2}{r_{i,j}}, 
\label{1}
\end{equation}
where $m_e$, $\bm{p}_i$, $r_i=|\bm{r}_i|$, and $r_{i,j}=|\bm{r}_i-\bm{r}_j|$ represent 
electron mass, momentum of $i$-th electron, distance in between a position of 
$i$-th electron and the origin, and relative distance in between $\bm{r}_i$ and $\bm{r}_j$.  
$v_{\rm core}$($r_i$) is a potential of atom and valence electrons (central force field 
approximation), 
$v_{\rm CF}$($r_i$) is a crystal field, 
$v_{\rm CF}(r_i)=\sum_{k=1}^6 \frac{Ze^2}{|\bm{R}_k-\bm{r}_i|}$, and the last term represents 
electron interaction between d electrons. 
When $n=1$, the fourth term vanishes which results in 
well known d$_{x^2-y^2}$ (we denote this as $\phi_v$ or $v$), 
d$_{3z^2-r^2}$ ($\phi_u$ or $u$), d$_{xy}$ ($\phi_\zeta$ or $\zeta$), 
d$_{yz}$ ($\phi_\xi$ or $\xi$), and d$_{zx}$ ($\phi_\eta$ or $\eta$) wave functions are obtained. 
Note that these wavefunctions are real functions. 
For $v_{\rm core}(r_i)=-\frac{e^2}{r_i}$, energy gap in between e$_{\rm g}$ and t$_{\rm 2g}$ 
orbitals is evaluated as 10$Dq$, where $D=\frac{35Ze}{4a^5}$ and $q=\frac{2e}{105}\langle r^4 \rangle$ 
(average of $r^4$, $\langle r^4 \rangle \equiv \int |R_{nd}|^2 r^4 r^2 dr$, $R_{\rm nd}$ is 
radial wavefunction of $nd$ states.) \cite{figgis}. 
Parameters $Z$ and $a$ can tune the energy gap. In this sense, $Z$ and $a$ can 
express material's characteristics. 
Thus, under octahedral coordination, 5-fold d orbitals split into 3-fold t$_{\rm 2g}$ 
and 2-fold e$_{\rm g}$ orbitals.  

When $n>1$, electron 
correlation should be taken into account. Tanabe and Sugano have constructed solutions for 
$n$ d-electrons ($1\leqq n \leqq 9$) in the crystal field (strong crystal field limit) \cite{tanabe1,tanabe2}. 
As a result, the many-electron state is found to be expressed as a linear combination of 
Slater determinants including $\phi_u$(or $u$), $\phi_v$($v$), 
$\phi_\zeta$($\zeta$), $\phi_\xi$($\xi$), $\phi_\eta$($\eta$) (spin up), 
$\bar{\phi}_u$(or $\bar{u}$), $\bar{\phi}_v$($\bar{v}$), 
$\bar{\phi}_\zeta$($\bar{\zeta}$), $\bar{\phi}_\xi$($\bar{\xi}$), $\bar{\phi}_\eta$($\bar{\eta}$) 
(spin down). 
Write the many-electron state as $\Phi_n$(t$_2^p$e$^q$:$^{2S+1}\Gamma$$M_{s}$$\gamma$), 
where $p$ and $q$ ($n=p+q$) represent a number of electrons at t$_2$ orbitals ($\zeta$, $\eta$, $\xi$) 
and e orbitals ($u$, $v$), $^{2S+1}\Gamma$ represents multiplet term, $M_{s}$ is an eigenvalue of 
total spin angular momentum ($S_z$), and $\gamma$ represents ground function of irreducible 
representation $\Gamma$ (e.g. $\Gamma=T_2$, $\gamma=\xi, \eta, \zeta$). 
As an example, we briefly treat $n=2$. A number of cases for (t$_2$)$^2$ configuration is 15 ($=_6$C$_2$). 
Thus, we can provide 15 Slater determinant as $|\xi \eta|$, $|\bar{\xi} \bar{\eta}|$, $\cdots$, 
and $|\xi\bar{\xi}|$.
Then, using a linear combination ($\Phi_2$) of these 15 Slater determinants, 
$\braket{\Phi_2|\frac{e^2}{r_{i,j}}|\Phi_2}$ can be diagonalized, then we obtain 
$\Phi_2 =\Phi_2$(t$_2^2$:$^{2S+1}\Gamma$$M_{s}$$\gamma$). This solution 
is further characterized by multiplet term which represents symmetry of the many-electron states 
under octahedral ligand field. 
For example, $\Phi_2$(t$_2^2$:$^{3}T_1$$M_{s}=1$$\gamma$)$=$
$|\xi \eta|$, $|\eta \zeta|$, and $|\zeta \xi|$. 
$\Phi_2$(t$_2^2$:$^{3}T_1$$M_{s}=0$$\gamma$)$=$$\frac{1}{\sqrt{2}}$
\{$|\xi \bar{\eta}|$-$|\eta \bar{\xi}|$\}, $\frac{1}{\sqrt{2}}$
\{$|\eta \bar{\zeta}|$-$|\zeta \bar{\eta}|$\}, and $\frac{1}{\sqrt{2}}$
\{$|\zeta \bar{\xi}|$-$|\xi \bar{\zeta}|$\}. 
$\Phi_2$(t$_2^2$:$^{3}T_1$$M_{s}=-1$$\gamma$)$=$
$|\bar{\xi} \bar{\eta}|$, $|\bar{\eta} \bar{\zeta}|$, and $|\bar{\zeta} \bar{\xi}|$. 
These 9 functions are energetically degenerated. A degeneracy of the multiplet term 
$^{2S+1}\Gamma$ is a product of spin multiplicity ($2S+1$) and $\Gamma_0$. 
$\Gamma_0$ is a dimensional number of irreducible representation 
(1 for A(B), 2 for E, and 3 for T). Thus, a degeneracy of $^3T_1$ is expressed as 
$3 \times 3 =9$.

Next, let's see briefly multiplet theory of atom \cite{griffith}. 
Many-electron state of a free ion is expressed using term symbol $^{2S+1}L$, where $L$ is total 
orbital angular momentum ($L=S, P, D, F, G, \cdots$ corresponds to $0, 1, 2, 3, 4, \cdots$, respectively.). 
A degeneracy of the many-electron state is expressed as ($2S+1$)($2L+1$).
For example, d$^5$ has 252-fold degeneracies ($=_{10}$C$_5$), however, under spherical coulomb potential, 
it splits and the ground term becomes to $^6$S. Thus, the degeneracy decreases from 252 to 6. 
Under weak crystal field, this ground states remain alive and expressed as $^6$A$_{\rm 1g}$. 
As shown above, for strong crystal field limit, we see how to construct many-electron states by 
Tanabe and Sugano. 
By connecting weak-crystal field limit \cite{griffith}, 
Tanabe and Sugano constructed Tanabe-Sugano diagram \cite{tanabe1,tanabe2}. 
Each states in the diagrams are labelled by term symbols, 
which shows symmetry of the many-electron states under ligand field and its degeneracies of each states. 
Thus, Tanabe-Sugano diagram gives information on the degeneracies of d electrons wavefunctions 
in octahedral ligand field. 
The number of the degeneracy can be read from the term symbol. 
Table I lists ground multiplet term and degeneracy ($\Gamma$) that 
the multiplet term represents for $0 \leqq n \leqq 10$. 
Note that the intermediate spin states are excited states in Tanabe-Sugano diagram. 
In our theory, the degeneracy of the ground multiplet term plays an important role. 
Ligand field theory succeeded in explaining several physical properties such as 
thermochemical properties (hydration enthalpies) and geometric distortions (Jahn-Teller distortion and 
spinel structures), various spectroscopies of transition 
metal coordination complexes, in particular optical spectra (colors), magnetic properties 
(spin-orbit-coupling related magnetism) and so on \cite{griffith,figgis}.

\subsection{Electronic structure of one-dimensional $N$-clusters chain model} 

\begin{table*}[t]
\centering
\begin{tabular}{lclcc}
\multicolumn{1}{c}{d$^n$} & \multicolumn{1}{c}{electronic configuration} & spin state &\multicolumn{1}{c}{ground multiplet term ($^{2S+1}\Gamma$)}& \multicolumn{1}{c}{degeneracy of $^{2S+1}\Gamma$ ($\Gamma)$)} \\
\hline
\hline
d$^0$ & (t$_{\rm 2g}$)$^0$(e$_{\rm g}$)$^0$& - &$\mathrm{^{1}A_{1g}}$&1\\
d$^1$ & (t$_{\rm 2g}$)$^1$(e$_{\rm g}$)$^0$& - &$\mathrm{^{2}T_{2g}}$&6\\
d$^2$ & (t$_{\rm 2g}$)$^2$(e$_{\rm g}$)$^0$& - &$\mathrm{^{3}T_{1g}}$&9\\
d$^3$ & (t$_{\rm 2g}$)$^3$(e$_{\rm g}$)$^0$& - &$\mathrm{^{4}A_{2g}}$&4\\
d$^4$ & (t$_{\rm 2g}$)$^3$(e$_{\rm g}$)$^1$& high spin &$\mathrm{^{5}E_{g}}$&10\\
         & (t$_{\rm 2g}$)$^4$(e$_{\rm g}$)$^0$& low spin &$\mathrm{^{3}T_{1g}}$&9\\
d$^5$ & (t$_{\rm 2g}$)$^3$(e$_{\rm g}$)$^2$& high spin &$\mathrm{^{6}A_{1g}}$&6\\
         & (t$_{\rm 2g}$)$^4$(e$_{\rm g}$)$^1$& $^*$intermediate spin &$^*$$\mathrm{^{4}T_{1g}}$&12\\
         & (t$_{\rm 2g}$)$^5$(e$_{\rm g}$)$^0$& low spin &$\mathrm{^{2}T_{2g}}$&6\\
d$^6$ & (t$_{\rm 2g}$)$^4$(e$_{\rm g}$)$^2$& high spin &$\mathrm{^{5}T_{2g}}$&15\\
         & (t$_{\rm 2g}$)$^5$(e$_{\rm g}$)$^1$& $^*$intermediate spin &$^*$$\mathrm{^{3}T_{1g}}$&9\\
         & (t$_{\rm 2g}$)$^6$(e$_{\rm g}$)$^0$& low spin &$\mathrm{^{1}A_{1g}}$&1\\
d$^7$ & (t$_{\rm 2g}$)$^5$(e$_{\rm g}$)$^2$& high spin &$\mathrm{^{4}T_{1g}}$&12\\
         & (t$_{\rm 2g}$)$^6$(e$_{\rm g}$)$^1$& low spin &$\mathrm{^{2}E_{g}}$&4\\
d$^8$ & (t$_{\rm 2g}$)$^6$(e$_{\rm g}$)$^2$& - &$\mathrm{^{3}A_{2g}}$&3\\
d$^9$ & (t$_{\rm 2g}$)$^6$(e$_{\rm g}$)$^3$& - &$\mathrm{^{2}E_{g}}$&4\\
d$^{10}$ & (t$_{\rm 2g}$)$^6$(e$_{\rm g}$)$^4$& - &$\mathrm{^{1}A_{1g}}$&1\\
\hline
\end{tabular}
\caption{\label{table1}Number of d electrons (d$_n$), electronic configuration, spin state, ground 
multiplet term ($^{2S+1}\Gamma$), 
and degeneracy of $^{2S+1}\Gamma$ ($\Gamma$). The ground terms were referred from Ref\cite{tanabe2}.$^*:$
these intermediate states are excited states in Tanabe-Sugano diagram. }
\end{table*}

\begin{figure*}[t]
\centering
\includegraphics[width=140mm,clip]{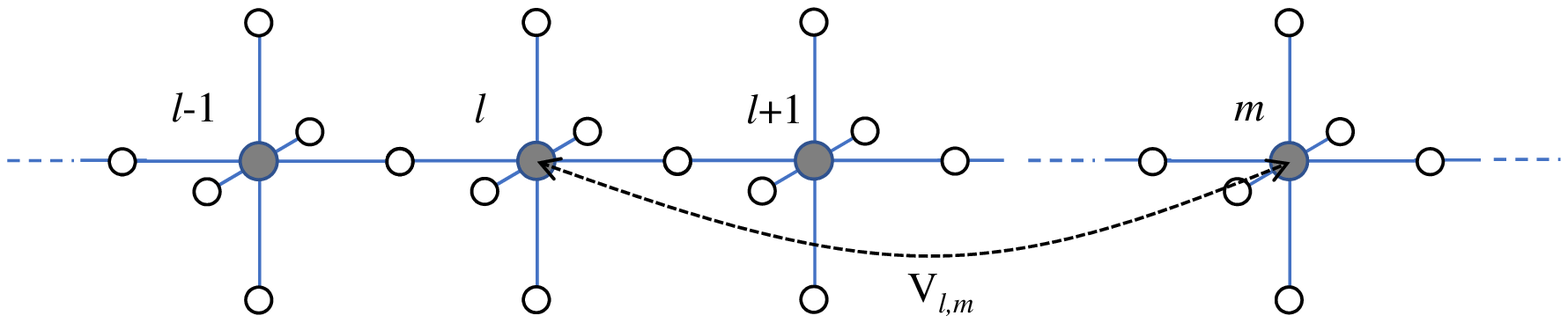}
\caption{(Color online) Schematic figure of periodically aligned one-dimensional N-clusters chain. 
The broken arrow represents inter-cluster interaction in between $l$-th and $m$-th clusters (V$_{l, m}$).
}
\label{f2}
\end{figure*} 

As a next step, let's construct one-dimensional $N$-clusters chain model. 
($N:$ a number of clusters)
We assume tight binding approximation of the clusters. 
The Hamiltonian of the clusters chain model is described as 
\begin{equation}
H_N= \sum_{l=1}^N H_{n_{l}l} + \sum_{l\neq m} V_{l,m}, 
\label{2}
\end{equation}
where $H_{n_{l}l}$ represent the Hamiltonian of the $l$-th cluster (equivalent to Eq. \ref{1}), 
$n_l$ represents a number of d electrons at $l$-th cluster, and  
inter-cluster interaction in between $l$-th and $m$-th clusters is introduced as $V_{l,m}$ (see Fig. \ref{f2}). 
The many-electron wave function of the $N$-clusters chain is 
constructed by a linear combination of one of the 
degenerated wave functions of the $l$-th cluster 
[$\Phi_{n_l l}$(t$_2^p$e$^q$:$^{2S+1}\Gamma$$M_{s}$$\gamma$)] as 
\begin{equation}
\Psi_N= \sum_{l=1}^N  c_l \Phi_{n_l l}(t_2^p e^q:^{2S+1}\Gamma M_{s} \gamma),  
\label{3}
\end{equation}
where $c_l$ is normalized constant. 
Since both $H_N$ and $\Psi_N$ are real function, the matrix elements are expressed as 
real symmetric matrix,  
\begin{multline}
\Braket{\Psi_N|H_N|\Psi_N} = \\
							\prescript{t\!}{}{
							\begin{pmatrix}
							\vdots \\
							\vdots \\
							c_l \\
							\vdots \\
							\vdots \\
							c_m \\
							\vdots \\
							\end{pmatrix}}
							\begin{pmatrix}
							\ddots &&&&&& \\
							&E_{l-1}&&&&& \\
							\dots&t_{l-1, l}&E_{l}& t_{l, l+1} &\dots&t_{l,m}& \dots \\
							&&\vdots& E_{l+1} &&& \\
							&&\vdots&&\ddots&& \\
							&&t_{m,l}&&&\ddots& \\
							&&\vdots&&&& \ddots \\
							\end{pmatrix}    
							\begin{pmatrix}
							\vdots \\
							\vdots \\
							c_l \\
							\vdots \\
							\vdots \\
							c_m \\
							\vdots \\
							\end{pmatrix},
\label{4}
\end{multline}
where $E_l$ is one cluster energy $\braket{\Phi_{n_l l}(t_2^p e^q:^{2S+1}\Gamma M_{s} \gamma|H_{nl}|\Phi_{n_l l}(t_2^p e^q:^{2S+1}\Gamma M_{s} \gamma}$, 
and 
$t_{l,m}$ is inter-cluster interaction energy 
$\braket{\Phi_{n_l l}(t_2^p e^q:^{2S+1}\Gamma M_{s} \gamma|V_{lm}|\Phi_{n_m m}(t_2^p e^q:^{2S+1}\Gamma M_{s} \gamma}$. 
$t_{l,m} = t_{m,l}$ is trivial as $V_{l,m}$ is two-body coulomb interaction. 
Thus, using a proper orthogonal matrix, this matrix can be exactly diagonalized, then 
eigenvalues and eigenstates of the N-clusters chain are exactly determined. 

Now, let's apply this general discussion to M$^{3+}$/M$^{4+}$ mixed-valent system. 
We set $E_l=E_3$ or $E_4$, where they are energy of a cluster with M$^{3+}$ or M$^{4+}$. 
A number of M$^{4+}$ clusters is $M$, and a number of M$^{3+}$ clusters is $N-M$. 
Under this condition, we take into account the nearest neighbor inter-cluster interaction, 
namely $t_{l-1,l}\neq 0$, $t_{l,l+1}\neq 0$, and $t_{l,m}= 0$ ($m\neq l-1$ or $l+1$).  
Total energy $E$(d$^n$) of the MX$_6$ cluster with d$^n$ is $E$(d$^n$)$=E_0 + n\epsilon_d^0 + 
\frac{n(n-1)}{2}\bar{U}$, where $E_0$ is total energy of d$^0$ cluster, 
$\epsilon_d^0$ is one-electron energy, and $\bar{U}$ is average of coulomb and exchange 
energies in between two electrons. Thus, $E$(d$^n$)$-$$E$(d$^{n-1}$) becomes 
$\epsilon_d^0+\bar{U}(n-1)$. Then, the $E_3 - E_4$ value can be regarded as Hubbard $U$. 
Thus, this approximation is regarded as Hubbard-like model. 
Then, the narrow band-like feature with the width of $W \sim t_{l-1,l}, t_{l,l+1}$ 
will be formed around $E_3$ and $E_4$. Using small inter-cluster interaction, 
thus, $\frac{U}{W} \gg 1$ can be realized. 

Here, we briefly discuss $t_{l-1,l}$ and $t_{l,l+1}$ from microscopic orbital point of view. 
The values of the hopping integrals in between s, p, and d orbitals for ($l', m', n'$) direction were already 
calculated as Slater-Koster parameter in Table I of Ref. \cite{slater}. According to Slater and Koster, 
the hopping integral in the high-symmetric one-dimensional model [two centers are connected along 
the directions of (1, 0, 0), (0, 1, 0), (0, 0, 1)] automatically becomes zero for several orbital cases. 
Thus, to avoid the zero hopping integral, tilted zigzag chain structure as seen in GdFeO$_3$-type 
structure is effective. Then, all the hopping in between d orbitals become possible, 
which means that electron can move from one edge to another edge through the 
crystal with energy, although the value of the hopping integral depends on the orbital. 
Then, if one uses a condition of $W \ll k_{\rm B}T \ll U$ (highly flat-band situation) 
where $k_{B}$ and $T$ represent Boltzmann constant and temperature, 
all the hopping can almost equally occur with use of the help of the thermal energy much 
larger than the band width. Thus, local degeneracy in the MX$_6$ octahedron can be kept 
even if such a flat band structure is constructed by the cluster chain. 
Since the orthorhombic GaFeO$_3$-type perovskite has three inequivalent directions ($a, b, c$), 
explicitly speaking, hopping integral along the one axis is slightly different from the ones along 
the other axes. However, the condition of $W \ll k_{\rm B}T \ll U$ can make the same local 
degeneracy along these axes. There are other structures composed by MX$_6$ octahedra. 
For example, edge-shared MX$_6$ octahedra can make a MX$_2$ layer with triangular lattice. 
For this case, t$_{\rm 2g}$-t$_{\rm 2g}$ hopping would be preferable compared with 
e$_{\rm g}$-e$_{\rm g}$ hopping. However, if such a material satisfies the condition 
of $W \ll k_{\rm B}T \ll U$, we can also use the local degeneracy to apply our theory to such a material. 

\subsection{Thermopower of the one-dimensional $N$-clusters chain model} 

Since the electronic structure of the N-clusters chain is exactly determined, 
now we can define energy flux ($J_q$) and current flux ($J$). Then, we can discuss thermopower. 
The expression of the thermopower is expressed as, 
\begin{equation}
S=-\frac{k_{\rm B}}{e}\left[\frac{1}{k_{\rm B}T}\left(\frac{J_q}{J}\right)_{\nabla T=0} - \frac{\mu}{k_{\rm B}T}\right],  
\label{5}
\end{equation}
where $\mu$ represents chemical potential \cite{marsh2}. When a material can be well described as 
band picture (mean-field approximation), 
the Boltzmann equation regime works well. Based on the band calculation and 
the Boltzmann equation, Singh reproduced a large thermopower ($\sim$ 110 $\mu$V/K) of 
NaCo$_2$O$_4$ at 300 K \cite{singh}. Even if band picture does not work well (ex. correlated 
hopping conduction), 
by carefully taken into account interactions in a Hamiltonian, Kubo-Luttinger formalism \cite{kubo,luttinger}
works well. Recently, Matsuura {\it et al.} explained large thermopower of ($\sim$ 20 mV/K) at 
around 10 K taking into account electron-phonon interaction \cite{matsuura}. 
Thermopower of strongly correlated electron system including 3d transition metal perovskite oxide 
has been qualitatively explained based on single-band Hubbard model \cite{marsh1,marsh2, koshibae1}. 
And introduced degeneracies of spin and orbital plays an important role. To further consider the 
degrees of freedom, we evaluate the thermopower of the one-dimensional N-clusters chain. 

First, we consider the term $\frac{1}{k_B T}\left(\frac{J_q}{J}\right)$. $\left(\frac{J_q}{J}\right)$ 
is an averaged energy that a carrier carries. For our model, 
against an external field [such as electric field $E$ or temperature gradient ($\nabla T$)] as perturbation, 
most of the initial states excites within the band width $W$. So that we may evaluate 
$\frac{1}{k_B T}\left(\frac{J_q}{J}\right)$ as $\sim \frac{W}{k_B T}$.
Thus, for $W\ll k_{\rm B} T\ll U$, at the limit of $T \to \infty$, 
\begin{equation}
\lim_{T \to \infty} S= \frac{\mu}{eT}=-\frac{1}{e}\left(\frac{\partial s}{\partial N_{\rm e}}\right)_{E, V},  
\label{6}
\end{equation}
where, $s$, $N_{\rm e}$, $E$, and $V$ represents entropy, electron number, energy and volume 
in the N-clusters chain, respectively. 
We call this high-temperature limit value. 
According to the $t-J$ model by Koshibae and Maekawa \cite{koshibae2}, 
the high-temperature limit value is almost realized at around $\frac{k_{\rm B}T}{t} \sim 5$.
According to one-dimensional Hubbard model analysis, $S$ almost saturates 
at around $\frac{k_{\rm B}T}{t} \sim 1$ \cite{zemljic}. 
This is also consistent with the assumption of our method [$(\frac{J_q}{J})\sim W$)]. 

When $V_{l-1.l}$ and $V_{l,l+1} \sim W$, although all the eigenstates and eigenvalues of 
the N-clusters chain system 
are exactly determined, calculation of the entropy (counting of a number of degenerated 
eigenstates at total energy $E$) is highly complicated. 
Thus, we further consider the limit of $W \to 0$ (Namely $V_{l,m} \to 0$). 
Here, we would like to note the difference in between zero-$\frac{W}{k_{\rm B}T}$ limit and 
zero-$W$ limit. The zero-$\frac{W}{k_{\rm B}T}$ limit does not mean $W=0$. 
To evaluate an exact number of degenerated eigenstates at $E$, we further impose 
strong limitation on our theory. 

At the $W=0$ limit, the N-clusters chain system becomes simple, namely a 
periodically-aligned one-dimensional noninteracting clusters. Then, the 
wavefunction of the system is described as a direct product of the wavefunction of 
the $l$-th cluster  [$\Phi_{n_l l}$(t$_2^p$e$^q$:$^{2S+1}\Gamma$$M_{s}$$\gamma$)], 
\begin{equation}
\Psi_N= \prod_{l=1}^N \Phi_{n_l l}({\rm t}_2^p {\rm e}^q:^{2S+1}\Gamma M_{s}\gamma).  
\label{7}
\end{equation}
Total energy $E$ of the system is expressed as $E=(N-M)E_3+ME_4$.
A number of electrons $N_{\rm e}$ is expressed as $N_{\rm e}=(N-M)n_{3}+Mn_4$, 
where $n_{3}$ ($n_{4}$) represents a number of electrons in a cluster with $M^{3+}$ ($M^{4+}$). 
Since $n_{3}-n_{4}=1$, $N_{\rm e}$ becomes $N_{\rm e}=Nn_{3}-M$.
Then, a number ($g$) of degenerated eigenstates at $E$ is evaluated as 
\begin{equation}
g=\Gamma_3^{N-M}\Gamma_4^M \frac{N!}{M!(N-M)!}, 
\label{8}
\end{equation}
where $\Gamma_3$ ($\Gamma_4$) represents degeneracy of many-electron states 
in a M$^{3+}$ (M$^{4+}$) cluster at $E_3$ ($E_4$). 
Using Boltzmann principle of entropy $s=k_{\rm B}\ln g$, thermopower at the high-temperature limit 
becomes 
\begin{equation}
S= -\frac{k_{\rm B}}{e}\left(\frac{\partial \ln g}{\partial N_{\rm e}}\right)_{E, V}
=\frac{k_{\rm B}}{e}\left(\frac{\partial\ln  g}{\partial M}\right)_{E, V}. 
\label{9}
\end{equation}
By substituting Eq. \ref{8} to Eq. \ref{9} and using Stirling's approximation, 
\begin{equation}
S= -\frac{k_{\rm B}}{e}\ln\left(\frac{\Gamma_3}{\Gamma_4}\frac{x}{1-x}\right), 
\label{10}
\end{equation}
is obtained as a final formula where $x$ is defined as $\frac{M}{N}$. 

\subsection{Comparison with the extended Heikes formula} 

The extended Heikes formula expresses thermopower in d electron system \cite{doumerc,
marsh1,marsh2,koshibae1}. 
The total number of configurations for $t \ll k_{\rm B}T\ll U$ will be written as 
\begin{equation}
g=g_3^{N_{\rm A}-M}g_4^M \frac{N_A!}{M!(N_A-M)!}, 
\label{11}
\end{equation}
where $N_A$ is a system size, 
$M$ is the number of M$^{4+}$ ions. $g_3$ and $g_4$ are defined as number of electronic 
configurations 
of M$^{3+}$ and M$^{4+}$ ions (spin and orbital degrees of freedom). 
Substituting Eq. \ref{11} for Eq. \ref{9}, then, the thermopower is obtained as 
\begin{equation}
S=-\frac{k_{\rm B}}{e}\ln\left(\frac{g_3}{g_4}\frac{x}{1-x}\right), 
\label{12}
\end{equation}
where $x$ is a ratio of M$^{4+}$ ion to the system size 
$N_{\rm A}$ ($x=\frac{M}{N_{\rm A}}$) \cite{marsh1,marsh2,koshibae1}. 
Eqs. \ref{11} and \ref{12} are almost identical to Eqs. \ref{8} and \ref{10}. 

Now let us compare $g_i$ with $\Gamma_i$ ($i=3, 4$). 
According to Marsh, Parris, and Koshibae {\it et al.,} \cite{marsh1,marsh2,koshibae1}, 
a number of $n$ d-electrons configurations in e$_{\rm g}$ and t$_{\rm 2g}$ orbitals with use of spin 
multiplicity ($2S+1$) and orbital degeneracy is calculated using a number of cases. 
When t$_{\rm 2g}$ or e$_{\rm g}$ orbital is partially occupied, 
direct product of totally symmetric representation and 
irreducible representation becomes $^1A_{1\rm g}\times ^{2S+1}\Gamma = 
^{2S+1}\Gamma$. So that degeneracy of orbital can be dimension of $\Gamma$ 
($\Gamma=A$ or $B \to 1, E \to 2, T \to 3$). 
Thus, degeneracy of the ground multiplet term becomes equal to degeneracies of spin and orbital.  
However, if both e$_{\rm g}$ and t$_{\rm 2g}$ orbitals are partially occupied (e.g. excited state such as 
intermediate spin state of Co$^{3+}$), product of representations becomes 
$E_{\rm g}\times T_{2\rm g} = T_{1\rm g} + T_{2\rm g}$ in $O_h$ symmetry. 
Then, the number of the configuration differs from the degeneracies of spin and orbital 
[e.g. t$_{\rm 2g}^4$e$_{\rm g}^1$ ($S=\frac{3}{2}$)$\to$ spin multiplicity is 4, orbital degeneracy is 3 for 
t$_{\rm 2g}$, 2 for e$_{\rm g}$, thus degeneracies of spin and orbital becomes 24 \cite{koshibae1}.]
because coulomb interaction split 24 states into $12+12$ states depending on the symmetry of the many 
electron states. 
However, if these states are regarded as degenerated due to some ignored interaction in 
the crystal field approximation \cite{koshibae1}, 
$g_i$ becomes identical to $\Gamma_i$ for $0 \le n \le 10$. 

Next, we see a difference of the derivations. 
Marsh and Parris calculated chemical potential with use of ``grand canonical ensemble'' \cite{marsh1,marsh2}. 
Koshibae {\it et al.,} calculated the chemical potential with use of ``micro canonical ensemble'' \cite{koshibae1}. 
Since different ensembles lead the same expression, the extended Heikes formula 
is validly constructed based on thermodynamics and quantum statistical mechanics. 
Our formula is consistent with their results as shown above. 
We derived our expression of $S$ by considering micro canonical ensemble of 
exact many-electron states at total energy $E$. 
Since the extended Heikes formula and our formula are almost identical, these formulae 
are complementary.

Thanks to the comparison shown above, now we recognize an important feature.  
When thermopower of the metallic perovskite with magnetic interaction is 
discussed, the Heikes formula has an advantage to guess the degeneracy of the many electron states. 
Since it is highly difficult to exactly know the eigenstates and 
the eigenvalues of correlated metallic states ($U\ll k_{\rm B}T$), 
our method is not applicable. 
Thus, thermopower in correlated metallic state even seems to be reproduced by the 
spin and orbital degeneracies. 
$S$ at high temperature limit is entropy that a carrier carries. 
So that the experimental data at high temperatures generally gives an important information 
about the entangled entropy. The extended Heikes formula is also applicable for 
even frustrated state with triangular lattice. 

\section{Comparison with experiments and discussion}

\begin{figure}[b]
\begin{center}
\includegraphics[width=85mm,clip]{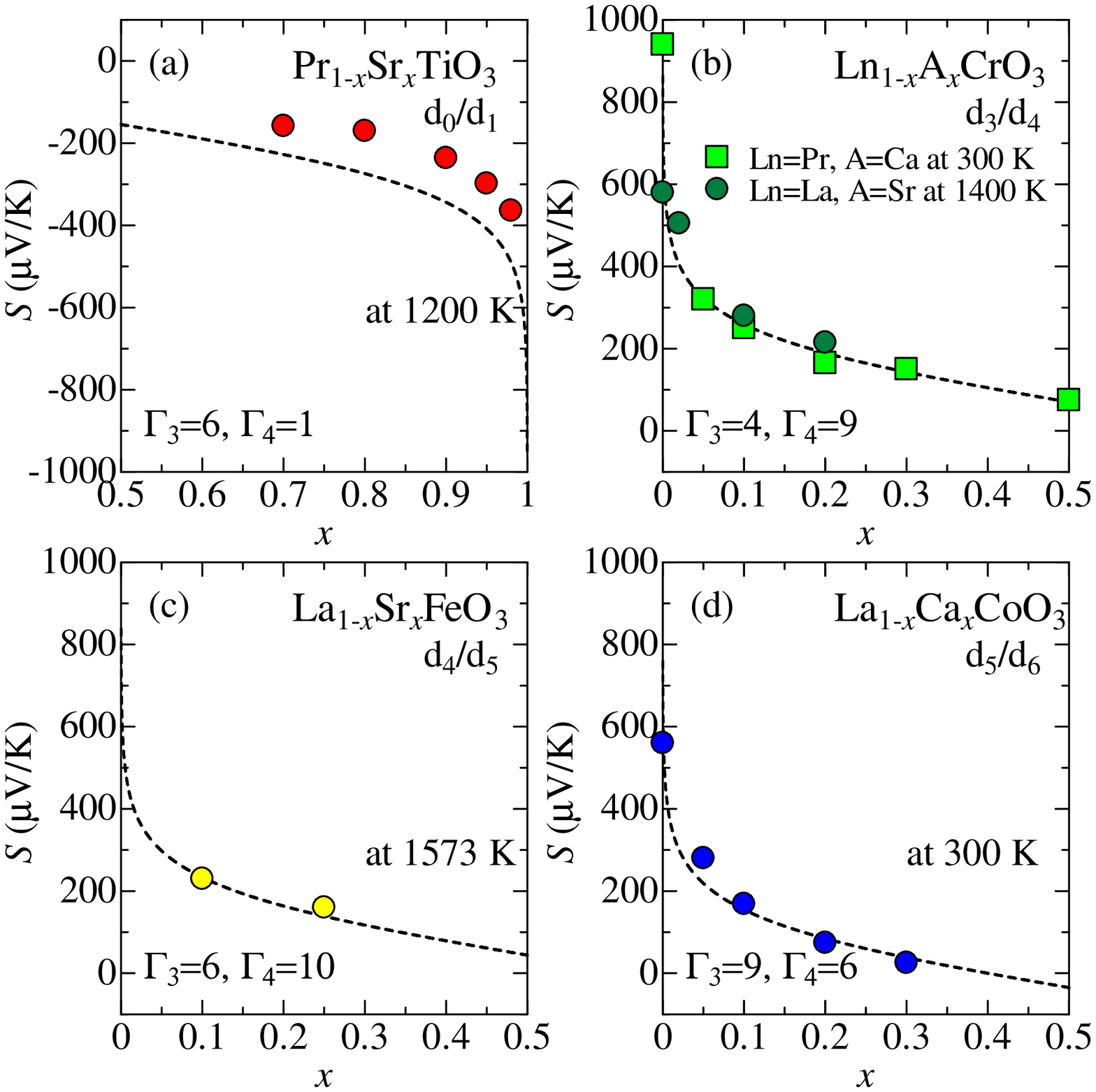}
\caption{(Color online) $x$ dependence of thermopower in 
(a) d$_0$/d$_1$-system Pr$_{1-x}$Sr$_x$TiO$_3$ at 1200 K\cite{kovalevsky}, 
(b) d$_3$/d$_4$-systems La$_{1-x}$Sr$_x$CrO$_3$ at 1400 K\cite{karim} and 
Pr$_{1-x}$Ca$_x$CrO$_3$ at 300 K\cite{pal}, 
(c) d$_4$/d$_5$-system La$_{1-x}$Sr$_x$FeO$_3$ at 1573 K\cite{mizusaki}, and 
(d) d$_5$/d$_6$-system La$_{1-x}$Ca$_x$CoO$_3$ at 300 K\cite{wang}. 
In each figure of (a)-(d), Eq. \ref{10} is drawn as a broken line. 
Numbers of $\Gamma_3$ and $\Gamma_4$ are displayed in the each figure.  
}
\label{f3}
\end{center}
\end{figure} 

As shown in the introduction, a condition, 
$W \ll k_{\rm B}T \ll U$ is realizing in 3d-transiton-metal perovskite with 
low $\tau$ (small $W$) \cite{imada}. 
At $V_{l,m}\sim 0$ ($W\ll k_{\rm B}T$), many of the system exhibit paramagnetic insulating state. 
Thus, 3d transition metal perovskite with insulating (semiconducting) conductivity 
due to hopping conduction and para magnetism will exhibit almost saturated thermopower at 
high temperatures. 
We investigate $x$ dependence of thermopower at high temperatures from 
previous works in which the material satisfies this condition. 
Now, let's compare experimental data with our expression. (Note that 
the proper data are not found after d$^7$ in the literature.)

\subsection{d$^0$/d$^1$ system}  

d$^0$/d$^1$ system corresponds to Ti$^{4+}$/Ti$^{3+}$ system. 
SrTiO$_3$ is a band insulator, and with La doping (electron doping), 
Sr$_{1-x}$La$_x$TiO$_3$ ($0.0 \leqslant x \leqslant 0.1$) exhibits paramagnetic metallic states. 
Thermopower of the system exhibits negatively large value, which is highly expected as a n-type 
thermoelectric material \cite{okuda}. 
With La$^{3+}$ $\to$ Pr$^{3+}$, Pr$_{1-x}$Sr$_x$TiO$_3$ exhibits insulating state 
possibly due to narrower $W$ than that of La$_{1-x}$Sr$_x$TiO$_3$. 

Figure \ref{f3}(a) shows $x$ dependence of thermopower in Pr$_{1-x}$Sr$_x$TiO$_3$ 
($0.7 \leqslant x \leqslant 0.98$) at 
1200 K \cite{kovalevsky}. With $\Gamma_3=6$ and $\Gamma_4=1$, Eq. \ref{10} is drawn 
as a broken line. 
The $x$ dependence of the thermopower is referred from the Fig. 5 in Ref. \cite{kovalevsky}. 
They still gradually increase even at 1200 K due to rather large $W$ of this system. 
However, the thermopower data seem to saturate at higher temperatures. So that we plot 
these data in Fig. \ref{f3}(a) with the theoretical curve. 

\subsection{d$^1$/d$^2$ system}

The filling controlled Mott transition system La$_{1-x}$Sr$_x$VO$_3$\cite{uchida}. 
is a typical d$^1$/d$^2$ system (V$^{4+}$/V$^{3+}$ system). 
According to the schematic metal-insulator diagram in Fig. 65 of Ref. \cite{imada}, 
this system has rather large $W$. So that 
thermopower of this system does not indicate its saturation 
(they exhibit strong $T$-dependence) 
even at high temperatures (1250 K). Thus, our expression can not treat this result. 
Combining the extended Heikes formula and dynamical mean field 
theory (DMFT) calculation on the single-band Hubbard model, 
M. Uchida {\it et al.,} indicate that the 
thermopower merge a value expected by the Heikes formula for $U\ll k_{\rm B}T$ limit at 
high temperatures.  

\subsection{d$^2$/d$^3$ system} 
La$_{1-x}$Sr$_x$CrO$_3$ is a typical d$^2$/d$^3$ system (Cr$^{4+}$/Cr$^{3+}$ system) 
which exhibits insulating $T$ dependence, and its thermopower almost saturates above 
1000 K \cite{karim}. Pr$_{1-x}$Ca$_x$CrO$_3$\cite{pal} has a narrower $W$ than that of 
La$_{1-x}$Sr$_x$CrO$_3$. The thermopower of this system almost saturate above 250 K. 
And paramagnetic insulating state is realized above 250 K. 
Pal {\it et al.,} has shown that the $x$ dependence of the thermopower is qualitatively 
explained by the extended Heikes formula. 
Marsh and Parris reproduced thermopower of La$_{1-x}$Sr$_x$CrO$_3$\cite{marsh1} 
system at 1400 K by using their theory with degeneracies of spin and orbital $\beta_0 =9$, and $\beta_1=4$. 
In Fig. \ref{f3}(b), we replotted the data of 
La$_{1-x}$Sr$_x$CrO$_3$ \cite{karim} and Pr$_{1-x}$Ca$_x$CrO$_3$ \cite{pal} with Eq. \ref{10}.

\subsection{d$^3$/d$^4$ system} 

La$_{1-x}$Sr$_x$MnO$_3$\cite{marsh2}, and La$_{1-x}$Ca$_x$MnO$_3$\cite{marsh2} are 
the typical model materials of d$^3$/d$^4$ system (Mn$^{4+}$/Mn$^{3+}$ system). 
From extensive research of these systems, a rich phase diagram is 
obtained. Charge ordering, ferromagnetism due to strong magnetic coupling, 
Jahn-Teller instability and phase separation cause the rich phases. 
Marsh and Parris explained the $x$ dependence of the thermopower of these systems using 
their theory \cite{marsh2}. 
It seems to us that in particular, $\Gamma_0 =1$ ($\Gamma_0: $ spin degeneracy) due 
to long-range magnetic coupling and 
$\Delta_{\rm JT}\ll k_{\rm B}T$, $U \ll k_{\rm B}T$ cause weak $x$ dependence of the 
thermopower below $x=0.2$. Palstra {\it et al.,} also reported $x$ dependence of the 
thermopower for La$_{1-x}$Ca$_x$MnO$_3$, and found almost $x$ independent values 
at 475 K \cite{palstra}. Kobayashi {\it et al.,} reported $x$ dependence of the 
thermopower of CaMn$_{3-x}$Cu$_x$Mn$_4$O$_{12}$ with narrower M-O-M bond angle 
($=142^{\circ}$), and found almost $x$ independent values 
at 1373 K \cite{kobayashi}. They exhibit insulating $T$ dependence of electrical conductivity 
possibly due to narrow band width $W$. However, Jahn-Teller instability and short- and long-range 
magnetic interaction seems to cause the small almost $x$ independent thermopower. 

\subsection{d$^4$/d$^5$ system} 

La$_{1-x}$Sr$_x$FeO$_3$\cite{mizusaki} is a typical d$^4$/d$^5$ system (Fe$^{4+}$/Fe$^{3+}$ system). 
La$_{1-x}$Sr$_x$FeO$_3$ ($x=0.1$ and 0.25) exhibit paramagnetic insulating state \cite{imada}. 
The thermopower almost saturate for $x=0.1$ above 1173 K. 
We plotted the data at 1573 K with $\Gamma_3=6$ and $\Gamma_4=10$ (for high spin) 
in Fig. \ref{f3}(c).

\subsection{d$^5$/d$^6$ system} 

LaCoO$_3$ is well-known as a spin-state crossover system \cite{goodenough}. With Sr doping, 
the system experiences MIT and shows ferromagnetic metallic state. 
Co-O-Co bond angle is smaller for $A=$ Ca than that for $A=$ Sr. 
La$_{1-x}$Sr$_x$CoO$_3$\cite{berggold,wang} is a typical d$^5$/d$^6$ system 
(Co$^{4+}$/Co$^{3+}$ system), and exhibits insulating temperature dependence 
of resistivity up to $x=0.2$ although it's $x=0.3$ for La$_{1-x}$Ca$_x$CoO$_3$. 
The thermopower seems to saturate around 300 K. (Note that at higher temperatures 
this system exhibits temperature induced MIT. Toward the MIT, $S$ decreases 
with $T$.) 
We replotted the thermopower data at 300 K in La$_{1-x}$Ca$_x$CoO$_3$\cite{wang} with $\Gamma_3=9$ 
and $\Gamma_4=6$ (intermediate state for Co$^{3+}$, low spin state for Co$^{4+}$) in Fig. \ref{f3}(d).

\subsection{general discussion}

As shown above, our formula well reproduces the 
thermopower of 3d transition metal perovskite oxides with small $W$ and 
large $U$ without both short- and long-range magnetic coupling. 
Although the extended Heikes formula 
seems to be applicable even for the correlated metallic state with magnetic interaction 
at a limit of $U \ll k_{\rm B}T$ (for V$^{4+}$/V$^{3+}$, and Mn$^{4+}$/Mn$^{3+}$ systems). 
In general, these conditions give weak $x$ dependence of the 
thermopower and rather small $S$ value due to small spin degree of freedom ($\sim 1$). 
Thus, entanglement in between other clusters works to decrease the thermopower. 
Jahn-Teller effect and spin orbit interaction can also be treated within our 
regime beyond the present evaluation. For example, La$_{1-x}$Ca$_x$MnO$_3$ has 
Jahn-Teller instability. The distortion of the MnO$_6$ octahedron induces 
$^5$E$_g$ $\to$ $^5$A$_{1g}$ $+$ $^5$B$_{1g}$, namely 10-fold degeneracies 
becomes 5-fold degeneracies with energetically stable state and 5-fold degeneracies with unstable state. 
This effect is treated as $\Gamma_3 =10 \to 5$ for d$^4$/d$^5$ system. 
As shown here, these effects break the degeneracy of the many electron states at any energy $E$, 
which gives smaller $S$ values than that the present one. Presently, our theory is applied to only 
the compounds with MX$_6$ octahedra. However, for example, if one uses local degeneracy 
of another cluster such as tetrahedron, triangular prism, and so on, 
one could discuss thermopower of the other structures with those clusters within the condition of 
$W \ll k_{\rm B}T \ll U$, which is beyond the present paper. 
In addition, we would like to note that our theory can treat d$^0$ to d$^{10}$ electrons, 
which includes other valences of M ion. 

\section{conclusion}

In conclusion, we widely investigate thermopower (at high temperatures) of 
3d transition metal perovskite oxides comparing the extended Heikes formula. 
We constructed an expression of the thermopower from many electron electronic states of 
MX$_6$ (M: transition metal, X: ligand element) octahedral clusters and 
reproduced $x$-dependence of the thermopower of several perovskites with $W \ll k_{\rm B}T \ll U$ 
using a degeneracy of many electron states that multiplet term represents at the total energy $E$. 
Comparison of our expression with the extended Heikes formula, complementarity of these 
formulae become clear. Thermopower in correlated metallic state with small 
$U$ and the magnetic coupling can be treated by the extended Heikes formula. 
This is highly efficient because without knowing the exact many electron states, we 
can evaluate thermopower with considering the inner degrees of freedom. 
Even in transport, we see an importance of crystal symmetry, which 
regulates a value of thermopower.

\acknowledgments

I would like to thank H. Kobayashi and S. Kobayashi for their supports.

\end{document}